\documentclass[superscriptaddress,nofootinbib,11pt,preprintnumbers,floatfix]{revtex4-2}
\usepackage[utf8]{inputenc}

\usepackage[dvipsnames]{xcolor}
\definecolor{red}{rgb}{0.9, 0,0}
\definecolor{cerulean}{rgb}{0., 0.42,0.9}
\definecolor{navy}{rgb}{0.05, 0.05,0.8}

\usepackage[colorlinks]{hyperref}
\hypersetup{
    colorlinks = true,
    citecolor  = red,
	linkcolor  = navy
}

\usepackage{slashed}
\usepackage{graphics}
\usepackage{amsmath}
\usepackage{amssymb}
\usepackage{latexsym}
\usepackage{mathtools}
\usepackage{dsfont}
\usepackage{hyperref}
\usepackage{amsfonts}
\usepackage{enumitem}
\usepackage{xstring} 
\usepackage{xspace} 
\usepackage[normalem]{ulem}
\usepackage{fontawesome}
\usepackage{subcaption}
\usepackage{aasmacros}

\newcommand{\Hz}{\,{\rm Hz}}

\newcommand{\pc}{\,{\rm pc}}

\newcommand{\yr}{\,{\rm yr}}

\def\vec#1{\mathbf{#1}}

\def\stat{\text{stat}}
\def\dyn{\text{dyn}}
\def\toa{\text{TOA}}
\def\red{\text{red}}
\def\dm{\text{dm}}

\newcommand{\comment}[1]{}

\newcommand\Tstrut{\rule{0pt}{3.5ex}}         
\newcommand\Bstrut{\rule[-2ex]{0pt}{0pt}}   

 \definecolor{lightblue}{rgb}{0.2,0.5,1}
 \hypersetup{colorlinks=true,
 linkcolor=lightblue,
 citecolor=lightblue,
 urlcolor=lightblue,
 linktocpage=lightblue,
 pdfproducer=lightblue}
\begin{document}

\title{Bayesian Forecasts for Dark Matter Substructure Searches with Mock Pulsar Timing Data}

\author{Vincent S. H. Lee}
\affiliation{Walter Burke Institute for Theoretical Physics, California Institute of Technology, Pasadena, CA}
\author{Stephen R. Taylor}
\affiliation{Department of Physics and Astronomy, Vanderbilt University, 2301 Vanderbilt Place, Nashville, TN 37235, USA}
\author{Tanner Trickle}
\affiliation{Walter Burke Institute for Theoretical Physics, California Institute of Technology, Pasadena, CA}
\author{Kathryn M. Zurek}
\affiliation{Walter Burke Institute for Theoretical Physics, California Institute of Technology, Pasadena, CA}

\preprint{CALT-TH-2021-016}

\begin{abstract}
    Dark matter substructure, such as primordial black holes (PBHs) and axion miniclusters, can induce phase shifts in pulsar timing arrays (PTAs) measurements due to gravitational effects. In order to gain a more realistic forecast for the detectability of such models of dark matter with PTAs, we propose a Bayesian inference framework to search for phase shifts generated by PBHs and perform the analysis on mock PTA data. For most PBH masses the constraints on the dark matter abundance agree with previous (frequentist) analyses (without mock data) to $\mathcal{O}(1)$ factors. This further motivates a dedicated search for PBHs (and dense small scale structures) in the mass range from $10^{-8}\,M_{\odot}$ to well above $10^2\,M_{\odot}$ with the Square Kilometer Array. Moreover, with a more optimistic set of timing parameters, future PTAs are predicted to constrain PBHs down to $10^{-11}\,M_{\odot}$. Lastly, we discuss the impact of backgrounds, such as Supermassive Black Hole Mergers, on detection prospects, suggesting a future program to separate a dark matter signal from other astrophysical sources.
\end{abstract}

\maketitle
\newpage
\tableofcontents
\newpage

\section{Introduction}
\label{sec:introduction}

Dark matter, despite being one of the most important components of standard cosmology, is not well-constrained on sub-galactic scales. The lack of observational constraints is problematic because many well-motivated models of dark matter predict unique structures on these small scales. For example, the Lambda Cold Dark Matter ($\Lambda$CDM) model with inflation produces a nearly scale invariant spectrum of adiabatic perturbations~\cite{Kolb1994, Dodelson2003} down to the free streaming scale corresponding to about $10^{-6}\,M_\odot$ for Weakly Interacting Massive Particle (WIMP) dark matter~\cite{Green_2005}. The QCD axion, if the Peccei-Quinn (PQ) symmetry~\cite{Peccei:1977hh} breaks after inflation, induces large isocurvature fluctuations on scales smaller than the QCD epoch horizon~\cite{Hogan:1988mp, Kolb:1993zz, Zurek_2007, Buschmann_2020, Arvanitaki_2020, Eggemeier_2020, xiao2021simulations}. Primordial black holes (PBHs) are generically formed by increasing the power of density fluctuations on small scales~\cite{Carr2020} which then collapse.

To date, general substructure constraints only extend down to mass scales $\sim 10\,M_\odot$, coming predominantly from gravitational microlensing of stars in the Large Magellanic Cloud and Andromeda~\cite{Alcock_2000, Tisserand_2007, Wyrzykowski_2011, Niikura_2019} as well as stars in the local neighborhood~\cite{Griest_2011, Griest_2014, Tilburg_2018}. Constraints on PBHs extend further down in mass due to their exceptionally high density. Non-evaporation from Hawking radiation requires $M \gtrsim 10^{-16}\, M_\odot$~\cite{Clark2016}, and microlensing currently constrains PBHs to be a subcomponent of dark matter for $M \gtrsim 10^{-10}\, M_\odot $~\cite{Croon2020, 2001, Niikura2019, Tisserand2007}.

It has been shown that pulsar timing arrays (PTAs) are potentially a powerful tool to search for dark matter substructure~\cite{Siegel:2007fz, Seto:2007kj, Baghram:2011is,Kashiyama_2012,Clark:2015sha,Schutz:2016khr,Dror:2019twh, Kashiyama:2018gsh,Ramani_2020,Lee:2020wfn} via \textit{Doppler} and \textit{Shapiro} effects. The Doppler effect is the change of observed pulsar frequency due to the acceleration of the pulsar as it is pulled by passing substructures gravitationally, while the Shapiro effect is a gravitational redshift effect due to the metric perturbations on the photon geodesic along the line of sight~\cite{Siegel:2007fz}. The signals can be further classified as \textit{static} (\textit{dynamic}) if the characteristic time scale of transiting dark matter, $\tau$, is much smaller (larger) than the pulsar observation time, $T$. A dynamic signal will be observed as a blip in the pulsar phase time series, whereas a static signal is observed as a long time scale perturbation. Generally static (dynamic) signals originate from heavier (lighter) dark matter, due to the smaller (larger) number density. 

In this paper, we develop techniques to detect signals from dark matter substructure that can be applied to real PTA data. Our purpose is to bridge the gap between the theoretically exhaustive analyses of Refs.~\cite{Dror:2019twh,Ramani_2020,Lee:2020wfn} and an application to real PTA data. We will focus our attention on monochromatic PBH dark matter since it is the simplest to study.  As noted previously~\cite{Dror:2019twh,Ramani_2020,Lee:2020wfn}, PTAs are sensitive to much less compact subhalos (such as axion miniclusters) than other lensing searches. To perform our analysis we use the software \texttt{enterprise} \citep{2019ascl.soft12015E} developed by the North American Nanohertz Observatory for Gravitational Waves (NANOGrav)~\cite{brazier2019nanograv}. \texttt{enterprise} utilizes a Bayesian inference framework to study how compatible pulsar phase models are with the measured data and noise sources. Since we are concerned not only with current PTAs but also future PTAs, such as the Square Kilometer Array (SKA)~\cite{Keane:2014vja}, we use realistic mock data which allows us to change certain PTA parameters, such as the number of pulsars or the observation time. Generally we find quite good agreement with the frequentist, signal-to-noise ratio (SNR) analysis performed previously in Refs.~\cite{Dror:2019twh,Ramani_2020,Lee:2020wfn} with the exception of PBHs with masses $10^{-4}\textup{--}10^{-2}\,M_{\odot}$ for the Shapiro search where the present constraint is closer to an order of magnitude weaker than the previous predicted constraints, mostly due to an approximation of the signal in the static regime necessary for carrying out the Bayesian analysis.\footnote{We leave for future work how to better approximate the Shapiro signal outside of the static regime. Note that the Shapiro dynamic signal rapidly becomes weak even for moderately lower halo concentration relative to a PBH \cite{Ramani_2020}, making the Shapiro dynamic search of limited utility for a broad range of dark matter models.}  Note that we restrict our work to the case of PBH dark matter, though, utilizing previous work \cite{Dror:2019twh,Ramani_2020,Lee:2020wfn}, our conclusions can be generalized to more diffuse substructures such as axion miniclusters.

To search for a dark matter signal with \texttt{enterprise}, a simple, parametrized form of the phase shift must be known. This precludes carrying out a fully general analysis, across static and dynamic signals which have dramatically different time series as discussed in Refs.~\cite{Dror:2019twh,Ramani_2020,Lee:2020wfn}.  We must break the analysis up into different regimes where a simple polynomial form of the signal dominates; we will nevertheless find that the three separate analyses we carry out with \texttt{enterprise} agree well across more than ten orders of magnitude in PBH mass with the frequentist approach carried out in Refs.~\cite{Dror:2019twh,Ramani_2020,Lee:2020wfn}. 

In particular, we parametrize the dark matter induced phase shift using an amplitude and at most one shape parameter. Schematically the detection pipeline consists of the following steps: 
\begin{enumerate}
    \item Search for the dark matter amplitude inside the PTA data.
    \item Compute the theoretical prediction of the dark matter amplitude.
    \item Compare the distributions of the amplitude from i) and ii) for consistency.
\end{enumerate}
The PTA data in i) are analyzed with \texttt{enterprise} while the theoretical predictions in ii) are computed numerically using the Monte Carlo (MC) simulations developed in Ref.~\cite{Lee:2020wfn}, which produces the probability distribution of the signal amplitudes.

Most importantly, we find that the leading order difference between the previous theoretical analyses and what can be realistically concluded with future PTA data will depend on how well a gravitational wave background (GWB) is separated and mitigated from a dark matter signal. Signals from GWBs are of primary interest for the PTA community. For instance, NANOGrav recently reported~\cite{Arzoumanian:2020vkk} strong evidence for a common-spectrum low-frequency stochastic process that is consistent with the characteristic strain spectrum from supermassive black hole binaries (SMBHBs~\cite{Sesana_2004, Burke_Spolaor_2019}). If handled naively over the entire frequency range of the data, and with no spatial correlation information included, we find that it will swamp a dark matter signal. While beyond the scope of this work, separating today's signals (such as the GWB) from signals of future interest (such as dark matter) will be crucial for the future science program of PTAs. The present work strongly motivates a focus on this type of background mitigation. 

The outline of the paper is as follows. In Sec.~\ref{sec:signal} we describe the form of the PTA signal injected by dark matter substructure, paying close attention to describing the needed approximations. In Sec.~\ref{sec:detection} we perform the Bayesian analysis with \texttt{enterprise} to derive the posterior distribution of dark matter amplitude in PTA data, and detail how to detect, or constrain, dark matter with this data. In Sec.~\ref{sec:mock} we apply these detection techniques to mock data and compare the constraints with our previous sensitivity projections in~\cite{Lee:2020wfn} and~\cite{Ramani_2020}. Finally, in Sec.~\ref{sec:conclusions} we conclude.

\section{Dark Matter Signals}
\label{sec:signal}

Pulsars are excellent tools for studying astrophysical phenomena because they are exceptionally stable clocks~\cite{1997A&A...326..924M}. Although the pulsar periods can fluctuate on shorter time scales, these fluctuations do not accumulate~\cite{1975ApJ...198..661H}. The intrinsic pulsar phase, $\phi(t)$, can then be modelled, to leading orders, by
\begin{align}
    \phi(t) = \phi_0 + \nu t+ \frac{1}{2}\dot{\nu}t^2\, ,
    \label{eq:timing_model}
\end{align}
where $\phi_0$ is the phase offset, $\nu$ is the pulsar frequency and $\dot{\nu}$ is its first derivative. This is called the \textit{timing model} of the pulsar. Since the second derivative of the pulsar frequency is small (typically $\Ddot{\nu}/\nu\lesssim 10^{-31} \,\Hz^2$~\cite{Liu_2018}), terms of order $\mathcal{O}(t^3)$ or higher are not included in the model. Any process that produces terms that are not in Eq.~\eqref{eq:timing_model} (e.g. a term $\propto t^3$) can be observed or constrained. In this section we focus on parametrizing additions to the pulsar phase due to a single dark matter subhalo.

The phase modification, $\delta \phi(t)$, induced by a dark matter subhalo can be written as
\begin{align}
    \delta\phi(t) = \int_0^t \delta \nu(t') dt' \, ,
    \label{eq:delta_phi_def}
\end{align}
where $\delta\nu$ is the induced frequency shift. The frequency shift due to Doppler and Shapiro effects were studied in Ref.~\cite{Dror:2019twh}, and are given by
\begin{align}
    \left(\frac{\delta \nu}{\nu}\right)_D &= \hat{\vec{d}} \cdot \int \nabla\Phi(\vec{r}, M) dt
    \label{eq:dnu_dop} \\
    \left(\frac{\delta \nu}{\nu}\right)_S &= -2 \int \vec{v} \cdot \nabla\Phi(\vec{r}, M) dz \, ,
    \label{eq:dnu_shap}
\end{align}
where $\hat{\vec{d}}$ is the unit vector pointing from Earth to the pulsar, $\Phi$ is the dark matter gravitational potential, $M$ and $\vec{v}$ are the mass and the velocity of the dark matter, respectively, and $z$ parameterizes the path that the photon travels from the pulsar to Earth. To further simplify these expressions, we write the position of the dark matter as $\vec{r}(t)=\vec{r}_0+\vec{v}t$ where $\vec{r}_0$ is the initial position.\footnote{We assume the dark matter subhalo travels in a straight line; a valid approximation when the orbital eccentricity $e \gg 1$. This requires $b\gg G(M+M_P)/v^2$~\cite{Jennings_2020} where $M_P$ is the pulsar mass. Since $M_P\approx1 \,M_{\odot}$~\cite{2016ARA&A..54..401O}, the impact parameter must satisfy $b\gg 10^{-8}\,\pc$, for $M\lesssim 1\,M_{\odot}$, which is indeed the case for the mass range considered in this work.} For the Shapiro signal, it is useful to define $\vec{r}_{\times}\equiv \vec{r}_0\times\hat{\vec{d}}$ and $\vec{v}_{\times}\equiv \vec{v}\times\hat{\vec{d}}$. Then the time for the dark matter to reach its point of closest approach is given by $t_{D,\,0}\equiv-\vec{r}_0\cdot\vec{v}/v^2$ and $t_{S,\,0}\equiv-\vec{r}_{\times}\cdot\vec{v}_{\times}/v_{\times}^2$, while the width of the signal is given by $\tau_D\equiv|\vec{r}_0\times\vec{v}|/v^2$ and $\tau_S\equiv|\vec{r}_{\times}\times\vec{v}_{\times}|/v_{\times}^2$. The impact parameter is $\vec{b}_D\equiv\vec{r}_0+\vec{v}t_{D,\,0}$ and $\vec{b}_S\equiv\hat{\vec{d}}\times(\vec{r}_{\times}+\vec{v}_{\times}t_{S,\,0})$. The explicit expressions for $\delta\phi(t)$ in the PBH limit, the main focus of this work, have been previously derived in Refs.~\cite{Dror:2019twh, Lee:2020wfn}, and are given by
\begin{align}
    \delta\phi_D(t) & = \frac{GM\nu}{v^2} \hat{\vec{d}}\cdot\left(\sqrt{1+x_D^2}\hat{\vec{b}}_D-\sinh^{-1}(x)\hat{\vec{v}}\right)
    \label{eq:dphi_dop} \\
    \delta\phi_S(t) & = 2GM\nu\log(1+x_S^2) \, ,
    \label{eq:dphi_shap}
\end{align}
where we define $x_D\equiv(t-t_{D,\,0})/\tau_D$ and $x_S\equiv(t-t_{S,\,0})/\tau_S$ as normalized time variables. We have also dropped all terms in Eq.~\eqref{eq:dphi_dop} and Eq.~\eqref{eq:dphi_shap} that are independent, linear or quadratic in time $t$ since they are completely degenerate with the timing model in Eq.~\eqref{eq:timing_model}, and hence unobservable. 

\subsection{Static and Dynamic Signals}
\label{subsec:stat_dyn_signals}

\texttt{enterprise} primarily uses a Markov Chain Monte Carlo (MCMC) to search over the parameter space in a signal model, which here is the dark matter signal. However such methods can become overwhelmed with too many variables, enhancing the search space dimensions, or variables degenerate in their effects, e.g. two variables describing the amplitude of a signal. This makes the expressions in Eq.~\eqref{eq:dphi_dop} and Eq.~\eqref{eq:dphi_shap} too cumbersome, and to facilitate the analysis, expressions of $\delta\phi(t)$ with fewer parameters are necessary. As discussed in Sec.~\ref{sec:introduction}, the signals can be further classified into static ($\tau\gg T$) and dynamic ($\tau \ll T$) signals. If the mass of the dark matter is large, the number density $n=\rho_{\dm}/M$ will be smaller, leading to a larger impact parameter. This translates to a large signal width since $\tau=b/v$. On the other hand, if the dark matter mass is small, the signal width $\tau$ will be small (precise definitions of `large' and `small' can be found in the discussion of the different length scales in Ref.~\cite{Ramani_2020}).

We start by discussing the Doppler effect. In the static limit, we can expand Eq.~\eqref{eq:dphi_dop} in a power series of $\tau/T$. Since all terms up to $O(t^2)$ are degenerate with the timing model, we can effectively parametrize the measurable signal as
\begin{align}
    \frac{\delta\phi_{D,\,\stat}(t)}{\nu} = \frac{A_{D,\,\stat}}{\yr^2} t^3 \, ,
    \label{eq:dphi_dop_stat}
\end{align}
where $A_{D,\,\stat}$ is a dimensionless parameter that characterizes the amplitude of the Doppler static signal and is given by
\begin{align}
    A_{D,\,\stat} = \yr^2 \frac{GM}{2v^2}\hat{\vec{d}}\cdot\left[\frac{t_{D,\,0}}{\tau_{D}^4}\frac{1}{(1+t_{D,\,0}^2/\tau_D^2)^{5/2}}\hat{\vec{b}}_D + \frac{1}{3\tau_D^3}\frac{1-2t_{D,\,0}^2/\tau_D^2}{(1+t_{D,\,0}^2/\tau_D^2)^{5/2}}\hat{\vec{v}}\right] \, .
    \label{eq:A_dop_stat}
\end{align}
We see that the static signal can be described by using only one parameter (i.e. $A_{D,\,\stat}$). In the dynamic limit, by observing that $\sqrt{1+x_D^2}\propto|t-t_{D,\,0}|$ when $\tau_D\ll T$, it is clear that up to a linear term in $x_D$, the phase shift is parametrized by
\begin{align}
    \frac{\delta\phi_{D,\,\dyn}(t)}{\nu} = A_{D,\,\dyn}(t-t_{D,\,0})\Theta(t-t_{D,\,0}) \, ,
    \label{eq:dphi_dop_dyn}
\end{align}
where $\Theta$ is the Heaviside step function and $A_{D,\,\dyn}$ characterizes the amplitude of the Doppler dynamic signal, which is given by\footnote{For simplicity, only the term $\propto \hat{\vec{b}}$ in Eq.~\eqref{eq:dphi_dop} is kept since it dominates over the term $\propto\hat{\vec{v}}$ in the dynamic limit.} 
\begin{align}
    A_{D,\,\dyn} = \frac{2GM}{v^2\tau_D}\hat{\vec{d}}\cdot\hat{\vec{b}}_D.
    \label{eq:A_dop_dyn}
\end{align}
We see that in contrast to the static signal, we need two different parameters ($A_{D,\,\dyn}$ and $t_{D,\,0}$) to fully describe the dynamic signal. 

We now turn our attention to the Shapiro effect. In a completely analogous way to the Doppler effect, we can parametrize the static Shapiro signal by a term $\propto t^3$
\begin{align}
    \frac{\delta\phi_{S,\,\stat}(t)}{\nu} = \frac{A_{S,\,\stat}}{\yr^2} t^3 \, ,
    \label{eq:dphi_shap_stat}
\end{align}
where $A_{S,\,\stat}$ is the amplitude of the signal given by
\begin{align}
    A_{S,\,\stat} = -\yr^2 \frac{4GM}{3}\frac{t_{S,\,0}}{\tau_S^4}\frac{3-t_{S,\,0}^2/\tau_S^2}{1+t_{S,\,0}^2/\tau_S^2} \, .
    \label{eq:A_shap_stat}
\end{align}
\begin{figure}[t!]
	\centering
	\includegraphics[width=\textwidth]{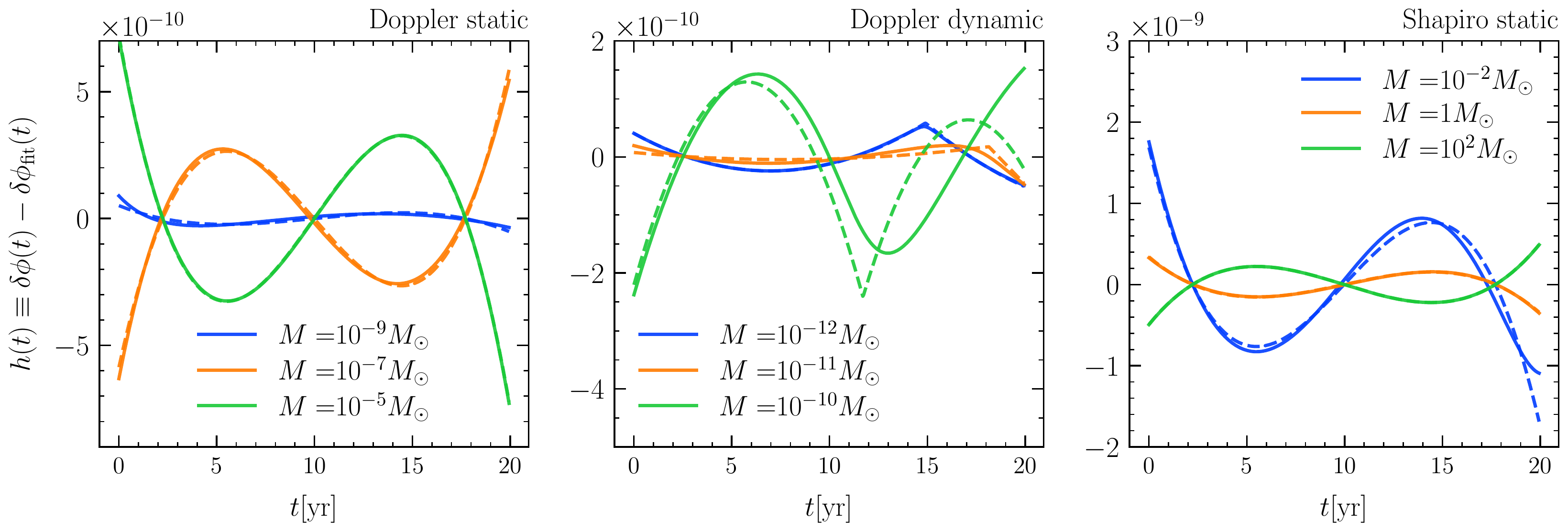}
	\caption{Comparison between the numerical and analytic subtracted timing residuals $h(t)$. The solid lines are generated from the MC with a single PBH using Eqs.~\eqref{eq:dphi_dop}-\eqref{eq:dphi_shap} while the dashed lines are computed using the analytic formulas in Eqs.~\eqref{eq:dphi_dop_stat}, \eqref{eq:dphi_dop_dyn} and \eqref{eq:dphi_shap_stat}. Both sets of signals are subtracted, meaning that the timing residual $\delta\phi(t)$ is first fitted to a second order polynomial in time. The fitted signal, $\delta\phi_{\mathrm{fit}}(t)$, is then subtracted from $\delta\phi(t)$.}
	\label{fig:spectrum}
\end{figure}
Parametrizing the Shapiro signal in the dynamic limit is tricky, since Eq.~\eqref{eq:dphi_shap} does not reduce to any simple expressions when $\tau_S \ll T$. On the other hand, in Refs.~\cite{Ramani_2020, Lee:2020wfn}, another type of signal known as the \textit{stochastic} signal is also considered. In the limit of extremely light substructure mass, a large number of events could collectively generate sizable signals. Moreover, the stochastic signal induces angular correlations between pulsars for the Earth term (similar to the GWB), which enhances the signal-to-noise ratio. However, parametrizing the stochastic signal is challenging (both in real and frequency space), and the Earth term analysis requires the construction of the likelihood function with non-square covariance matrices, which is computationally demanding. In light of these challenges, we do not search for these signals in this work and leave the analyses for future work.

In Fig.~\ref{fig:spectrum} we show some characteristic signal shapes of the timing residuals generated from the MC\footnote{Not to be confused with the MCMC introduced earlier in this subsection, which is a sampling scheme used to explore the parameter space and compute the posterior distribution.} after numerically fitting away all terms of order $\mathcal{O}(t^2)$ or less in the time series. For the static searches, we observe that the subtracted signals closely resemble a cubic polynomial in $t$, which justifies the $t^3$ parametrization that we have taken. For the Doppler dynamic case, the timing residuals have a rather abrupt turn at $t=t_{D,\,0}$, which matches with our prediction of the signal behaving like a step function in this limit. As shown in the figure, we find good agreement between the numerical results and the analytic approximations.

The important physical parameters are the dark matter mass, $M$, and mass fraction, $f_{\dm}\equiv \Omega / \Omega_{\text{dm}}$, where $\Omega_{\text{dm}}\equiv\rho_{\dm}/\rho_{\mathrm{crit}}$ is the local dark matter density parameter and $\Omega\equiv\rho/\rho_{\mathrm{crit}}$ is the local density parameter of the dark matter of interest (PBHs in this work). The relations between these parameters and the signal amplitudes $A$ are obtained using MC simulations described in Ref.~\cite{Lee:2020wfn}. While the MC simulations can generate signals from general dark matter subhalos, we focus on the PBH case here. We first randomly distribute PBHs with mass $M$, density $\rho_{\dm}f_{\dm}$ ($\rho_{\dm}=0.46\,\mathrm{GeV/cm^3}$~\cite{Sivertsson_2018}), and a Maxwell-Boltzmann velocity distribution with $v_0=325\,\mathrm{km/s}$, $v_{\mathrm{esc}}=600\,\mathrm{km/s}$ and isotropic angular dependence.\footnote{We have taken a dark matter velocity spread, $v_0$, higher than the often quoted value $v_0\approx 230\,\mathrm{km/s}$. This is to eliminate the velocity anisotropy due to the relative motion between the pulsar and the galactic rest frame. Since we do not expect such anisotropy to be observable, we ignore this effect and boost the distribution with a larger $v_0$ value.} The simulation volume is taken to be a sphere for the Doppler search and a cylinder with height $d$ for the Shapiro search, where $d$ is the distance between the pulsar and Earth. The center of the simulation volume is taken to be the position of the pulsar. The PBHs are then classified as dynamic if they satisfy $T-\tau>t_0>\tau$ and static otherwise~\cite{Dror:2019twh}. This condition ensures that the pulsar phase shift behaves approximately like Eq.~\eqref{eq:dphi_dop_dyn} for the dynamic PBHs. To compute $A_{\stat}$, we first evaluate the total pulsar phase shift (as a function of time) due to all the static PBHs using Eq.~\eqref{eq:dphi_dop} and Eq.~\eqref{eq:dphi_shap} for the Doppler and the Shapiro case respectively. Then we fit the phase shift to a cubic polynomial in time to extract the coefficient of the $t^3$ term, which gives us $A_{D,\,\stat}$ and $A_{S,\,\stat}$ in accordance with Eq.~\eqref{eq:dphi_dop_stat} and Eq.~\eqref{eq:dphi_shap_stat}. To compute $A_{D,\,\dyn}$, we use Eq.~\eqref{eq:dphi_dop_dyn} for the dynamic PBH that has the smallest $\tau_D$. Finally, we repeat the simulation for numerous realizations to obtain the conditional probability distributions $P(A_{D,\,\stat}|f_{\dm})$, $P(A_{D,\,\dyn}|f_{\dm})$ and $P(A_{S,\,\stat}|f_{\dm})$ for each choice of $M$.\footnote{We have suppressed the PBH mass $M$ inside the conditional probability for notational convenience in later sections. A larger $f_{\dm}$ implies a larger signal amplitude $A$, so it is conceptually more natural to draw upper limits on $f_{\dm}$ for each choice of $M$ instead of deriving the two dimensional posterior distribution for both parameters simultaneously.}

\section{Bayesian Analysis of Dark Matter Signals in PTAs}
\label{sec:detection}

We now develop the Bayesian framework for detecting dark matter subhalos with PTAs. For clarity, we will collectively refer the dark matter signal amplitudes for the different types of signals $A_{D,\,\stat}$, $A_{D,\,\dyn}$ and $A_{S,\,\stat}$ defined in Eq.~\eqref{eq:dphi_dop_stat}, Eq.~\eqref{eq:dphi_dop_dyn} and Eq.~\eqref{eq:dphi_shap_stat} as $A$.

\subsection{Noise Modeling and Likelihood}

Our modeling and analysis of PTA data closely follows Refs.~\cite{Arzoumanian_2016, 2015}, and we refer the reader to those papers for a full discussion of the PTA data model. We summarize several of the salient features here. Let $N_{\toa}$ be the number of pulsar times of arrival (TOAs). The timing residuals of a pulsar correspond to the raw TOA data with the best-fit timing model subtracted. By definition, any unmodeled phenomena or noise fluctuations should be encapsulated in the residuals, which we model as
\begin{align}
    \delta\vec{t} = M\vec{\epsilon} + F\vec{a} + \delta\vec{t}_{\dm} + \vec{n} \, .
    \label{eq:timing_residual_breakdown}
\end{align}
The matrix $M$ is the timing model design matrix corresponding to partial derivatives of the TOAs with respect to timing model parameters, and the vector $\vec{\epsilon}$ denote small linear parameter offsets. Together $M\vec{\epsilon}$ represents the inaccuracies in the subtraction in the timing model. 

The term $F\vec{a}$ represents a Fourier series of low-frequency (`red') timing deviations, where $F$ is an $N_{\toa}\times2N_{\text{modes}}$ matrix with alternating columns of sines and cosines in harmonics of the base frequency $1/T$, and $\vec{a}$ are the Fourier coefficients of each mode. Sources of pulsar intrinsic red noise include spin instability noise, secular pulse profile changes, and time-dependent dispersion measure variations~\cite{2013CQGra..30v4002C, Lam_2016, Jones_2017} (although the latter has a known dependence on the observed radio frequency). Inter-pulsar correlated red noise may derive from Roemer-delay errors when barycentering the pulse TOAs (inducing dipolar correlations) \citep{2016MNRAS.455.4339T}, long-timescale drifts in clock standards (inducing monopolar correlations) \citep{2016MNRAS.455.4339T}, and a stochastic GWB that is generated from a population of sources such as SMBHBs (inducing quadrupolar-dominated Hellings \& Downs correlations \citep{1983ApJ...265L..39H})~\citep[{\em e.g.},][and references therein]{2019A&ARv..27....5B}. We do not consider barycentering or clock errors here, nor do we leverage the Hellings \& Downs correlations between pulsars for the GWB; for the sake of computational convenience, the GWB is modeled as an uncorrelated common-spectrum red process amongst all pulsars, as in the NANOGrav 12.5yr Dataset analysis \citep{Arzoumanian:2020vkk}. 

Red noise of any source is modeled as a stationary Gaussian process with a power-law power spectral density of timing deviations:
\begin{align}
    P_{\red}(f) = \frac{A_{\red}^2}{12\pi^2}\left(\frac{f}{\yr^{-1}}\right)^{-\gamma_{\red}}\yr^3 \, ,
    \label{eq:red_noise_power_law}
\end{align}
where $A_{\red}$ and $\gamma_{\red}$ are the red noise amplitude and the spectral index respectively. For a GWB produced by a population of SMBHBs evolving solely through GW emission, $\gamma=13/3$ \citep{Phinney:2001di}. 

The term $\vec{n}$ denotes white noise that has equal power across all frequencies in the residual time series, and which is uncorrelated amongst pulsars. This noise is heteroscedastic with a per-TOA uncertainty dominated by the pulse template-fitting uncertainties. These uncertainties are then scaled. NANOGrav also computes many near-simultaneous sub-banded TOAs, producing white noise that is correlated across sub-bands, but uncorrelated in time. Once all of these effects are accounted for, the white noise covariance matrix has a block-diagonal structure in epoch blocks. 

The term $\delta\vec{t}_{\text{dm}}$ denotes a putative dark matter signal, which we model as a deterministic process. Grouping the timing model offsets and red noise together into the matrix-vector product $T\vec{b}_{\mathrm{lat}}$, we form model-dependent white noise residuals, $\vec{r}_{\mathrm{res}} = \delta\vec{t}-T\vec{b}_{\mathrm{lat}}-\delta\vec{t}_\dm$. The likelihood is then simply a Gaussian distribution in all the data with zero mean and a covariance matrix given by the modeled white-noise. However, we are typically not interested explicitly in the latent parameters $\vec{b}_{\mathrm{lat}}$, such that we analytically marginalize over these parameters with Gaussian priors described by the unbounded variance of the timing model offsets and the power spectral density (PSD) of the red noise. The resulting likelihood function is then 
\begin{align}
    p(\delta\vec{t}|\vec{\eta},\vec{\theta})=\frac{\exp(-\frac{1}{2}(\delta\vec{t}-\delta\vec{t}_\dm(\vec{\theta}))^T C(\vec{\eta})^{-1} (\delta\vec{t}-\delta\vec{t}_\dm(\vec{\theta})) )}{\sqrt{\text{det}(2\pi C(\vec{\eta}))}} \, . 
    \label{eq:likelihood}
\end{align}
where $\vec{\eta}$ are hyper-parameters describing the spectral models of the intrinsic pulsar red noise and GWB; $C$ is the model-dependent covariance matrix of white noise, red noise, and timing offsets; and $\vec{\theta}$ are parameters of the dark matter signal. This likelihood is constructed using the \texttt{enterprise} \citep{2019ascl.soft12015E} and \texttt{enterprise$\_$extensions} \citep{enterprise_ext} software packages, and the Bayesian posterior distributions of all parameters are sampled using MCMC techniques implemented with the \texttt{PTMCMCSampler} package \citep{2019ascl.soft12017E}. 
To compute the posterior distribution of the dark matter amplitude, we numerically marginalize the MCMC chain over all parameters except this amplitude. For the Doppler dynamic search, the time of arrival $t_0$ is also marginalized over. In every case, we obtain the posterior distribution of the dark matter amplitude $P(A|\delta \vec{t})$.

\subsection{Posterior Distribution of the Dark Matter Abundance}
\label{subsec:pos_dist_of_DM}

As stated in the previous section, the physical parameters that we are interested in are the dark matter mass $M$ and mass fraction, $f_{\dm} \equiv \Omega/\Omega_\dm $. This subsection describes the translation from the posterior distribution on the amplitude $A$, $P(A|f_{\dm}, M)$ to a statement on the dark matter abundance. 

\subsubsection{Single Pulsar}
\label{subsubsec:single_pulsar}

We begin with the simple case of a single pulsar, and fix the dark matter subhalo mass $M$ for the remainder of this subsection. Even for a fixed $f_{\dm}$, the amplitude $A$ is a random variable since both $\vec{r}_0$ and $\vec{v}$ are random variables. The conditional probability $P(A|f_{\dm})$ can be computed using the MC simulation described in Sec.~\ref{subsec:stat_dyn_signals}. The marginalized posterior distribution of $f_{\dm}$, given the measured data $\delta \vec{t}$, is
\begin{align}
    P(f_{\dm}|\delta \vec{t}) = \int_{-\infty}^{\infty}P(f_{\dm}|A)P(A|\delta \vec{t})dA \, .
    \label{eq:post_f}
\end{align}
Using Bayes' theorem, we can invert the conditional probability
\begin{align}
    P(f_{\dm}|A) = \frac{P(A|f_{\dm})P(f_{\dm})}{P(A)} \, ,
    \label{eq:bayes}
\end{align}
and assuming uniform priors on both $f_{\dm}$ and $A$, we can write
\begin{align}
    P(f_{\dm}|A) \propto P(A|f_{\dm}) \, . 
    \label{eq:prop}
\end{align}
Substituting Eq.~\eqref{eq:prop} into Eq.~\eqref{eq:post_f} gives
\begin{align}
    P(f_{\dm}|\delta \vec{t}) \propto \int_{-\infty}^{\infty}P(A|f_{\dm})P(A|\delta \vec{t})dA \, , 
    \label{eq:post_prop}
\end{align}
subjected to the normalization condition, $\int_{0}^{\infty} P(f_{\dm}|\delta \vec{t}) df_{\dm}=1$.

\subsubsection{Multiple Pulsars}

The above analysis is easily generalized to multiple pulsars. The marginalized posterior distribution of $f_{\dm}$ for multiple pulsars can be formulated in two non-equivalent, but equally valid, ways. First, we write the collection of the amplitude in each pulsar as $\vec{A}=(A_1, A_2, \cdots, A_{N_P})$ where $N_P$ is the number of pulsars. Then $P(f_{\dm}|\delta \vec{t})$ is given by
\begin{align}
    P(f_{\dm}|\delta \vec{t}) = \int_{-\infty}^{\infty}P(f_{\dm}|\vec{A})P(\vec{A}|\delta \vec{t})d^{N_P}A \, .
    \label{eq:post_f_multi}
\end{align}
Since all the pulsars are independent from each other, we can factorize the likelihood function and hence the joint posterior distribution of $\vec{A}$ (since $\vec{A}$ has a uniform prior) 
\begin{align}
    P(\vec{A}|\delta \vec{t}) = P(A_1|\delta \vec{t})P(A_2|\delta \vec{t})\cdots P(A_{N_P}|\delta \vec{t}) \, .
    \label{eq:A_fac}
\end{align}
Following the same steps in Sec.~\ref{subsubsec:single_pulsar}, the final expression of $P(f_{\dm}|\delta \vec{t})$, labelled `all' because it includes all the pulsars directly, is
\begin{align}
    P_{\mathrm{all}}(f_{\dm}|\delta \vec{t}) \propto \prod_{i=1}^{N_P} \int_{-\infty}^{\infty} P(A_i|f_{\dm})P(A_i|\delta \vec{t}) dA_i \, ,
    \label{eq:post_prop_multi}
\end{align}
which must also be normalized to one. We emphasize that since Eq.~\eqref{eq:post_prop_multi} is merely a product of $N_P$ integrals (instead of an $N_P$-dimensional integral), it is computationally inexpensive to evaluate.

Alternatively, instead of using the amplitudes from all pulsars, we can compute $P(f_{\dm}|\delta \vec{t})$ using only the pulsar with the maximum amplitude, labelled `max',
\begin{align}
    P_{\mathrm{max}}(f_{\dm}|\delta \vec{t}) \propto \int_{-\infty}^{\infty} P(A_{\max}|f_{\dm})P(A_{\max}|\delta \vec{t})dA_{\max} \, ,
    \label{eq:post_prop_multi_max}
\end{align}
where $A_{\max}\equiv \max_{i}A_i$. The upper limits placed on $f_{\dm}$ for these two different ways of formulating $P(f_{\dm}|\delta \vec{t})$ scale differently with $N_P$. If we use the amplitudes from all pulsars, it is clear from Eq.~\eqref{eq:post_prop_multi} that $P_{\mathrm{all}}(f_{\dm}|\delta \vec{t})$ we obtain from considering $N_P$ pulsars is effectively raising the single pulsar posterior by a factor of $N_P$ (up to normalization, assuming identical pulsars), which always results in a lower $90^\text{th}$ percentile on $f_{\dm}$. On the other hand, if we only consider the pulsar with the maximum amplitude, since $A_{\max}\geq A_i$ for all $i$, $P_{\mathrm{max}}(f_{\dm}|\delta \vec{t})$ will also be shifted to lower $f_{\dm}$. Hence we get a more stringent upper limit on $f_{\dm}$ with larger $N_P$ for both treatments, but they do not necessarily scale with the same power of $N_P$. Since both $\vec{A}$ and $A_{\max}$ are well defined statistical variables, we have the freedom to draw upper limits on $f_{\dm}$ using either of them (despite the fact that they give different results). These treatments can now be repeated for all choices of $M$ to obtain $P(f_{\dm}|\delta \vec{t})$ for each $M$.

\section{Mock Data}
\label{sec:mock}

To demonstrate the formalism developed in the previous sections we place the upper limits on the dark matter abundance in PBHs with standard mock pulsar data.

\subsection{Dataset}
The mock pulsars in our analyses originated from the International Pulsar Timing Array (IPTA) First Mock Data Challenge (MDC)~\cite{Verbiest_2016}. Using the python wrapper \texttt{libstempo}~\cite{2020ascl.soft02017V} to the pulsar timing package \texttt{TEMPO2}~\cite{Hobbs_2006, Edwards_2006}, we generate mock data from the MDC \texttt{.par} files with zero timing residuals (i.e. perfect fit of the timing model). We then prepare two sets of mock pulsars with the pulsar parameters consistent with the predicted parameters for future PTA experiments, which are summarized in Table~\ref{tab:pulsar_parameters}.

\begin{table*}[t]
  \begin{tabular*}{\textwidth}{c @{\extracolsep{\fill}} ccccc}
\hline
 & $N_P$ & $d\, [\mathrm{kpc}]$ & $T\, [\mathrm{yr}]$ & $\Delta t\, [\mathrm{week}]$ & $t_{\mathrm{rms}}\, [\mathrm{ns}]$   \\ 
\hline 
\hline
 SKA & 200 & 5 & 20 & 2 & 50 \\ 
 Optimistic & 1000 & 10 & 30 & 1 & 10 \\ 
 \hline
\end{tabular*}
\caption{\label{tab:pulsar_parameters} PTA parameters assumed when generating the mock pulsars. Here $N_P$ is the number of pulsars, $d$ is the pulsar-Earth distance, $T$ is the observation time, $\Delta t$ is the cadence and $t_{\mathrm{rms}}$ is the root-mean-square timing residuals.}
\end{table*} 

Then, for each set of mock pulsars, we inject noise into the timing residuals. To compare with our previous works~\cite{Ramani_2020, Lee:2020wfn}, our main result uses mock pulsars with only white noise injected. In addition, we also prepare a separate set of mock pulsars with both white noise and red noise injected. The spectral index of the red noise is chosen to be $\gamma_{\mathrm{red}}=13/3$, which is the theoretical prediction of a stochastic gravitational wave background (GWB) signal due to a population of inspiraling SMBHBs in circular orbits~\cite{Phinney:2001di}. To investigate the effects of red noise with different amplitudes, we carried out the analysis using mock data with $A_{\red}=10^{-17}$, $10^{-16}$, $10^{-15}$ and $10^{-14}$.

\subsection{Results}

To generate the posterior distribution of the dark matter signal amplitude, $A$, we closely follow the Bayesian inference procedure described in Ref.~\cite{Arzoumanian_2016} using the software \texttt{enterprise}~\cite{2019ascl.soft12015E}. We marginalize over all the timing model and noise parameters with an MCMC using the package \texttt{PTMCMCSampler}~\cite{2019ascl.soft12017E}. The time of closest approach $t_{D,\,0}$ for the Doppler dynamic search is also marginalized over. Since the signals we are interested in are pulsar independent, we carry out the analysis independently for each pulsar. 

\begin{figure}[t!]
    \centering
    \includegraphics[width=\textwidth]{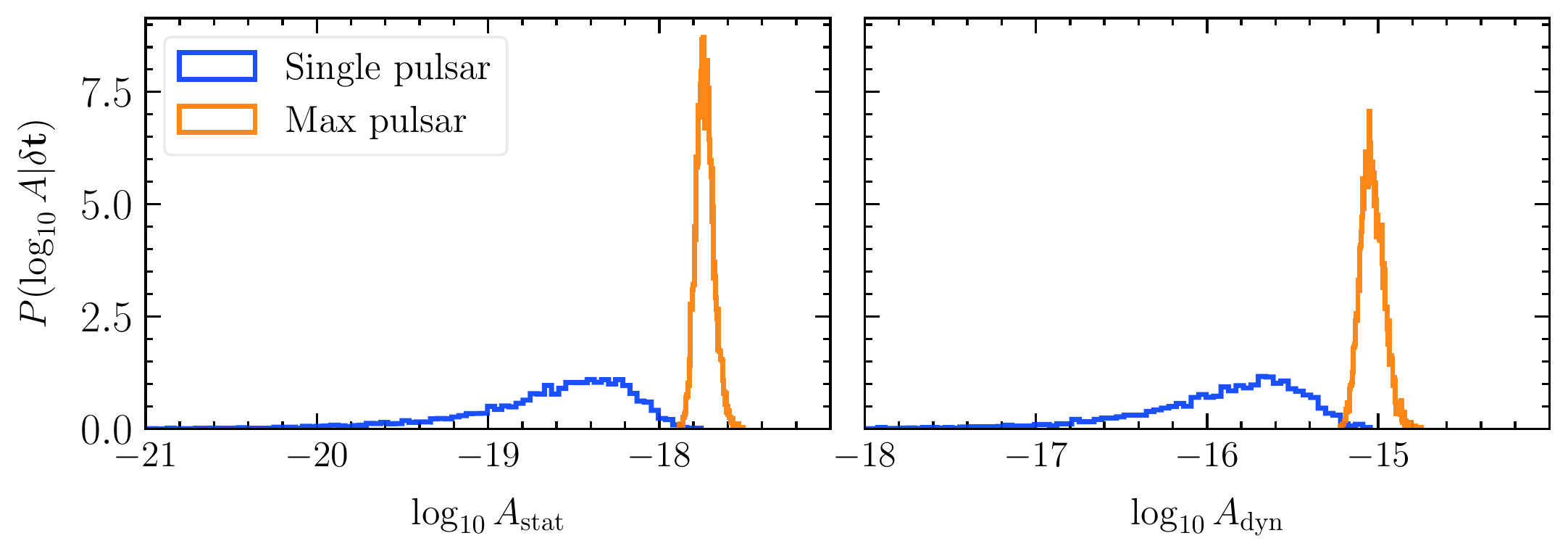}
    \caption{Posterior distribution of $\log_{10}A_{\stat}$ and $\log_{10}A_{\dyn}$ for mock pulsars with white noise only assuming SKA parameters. Both the single pulsar posterior and the posterior of the maximum amplitude across all $N_P=200$ pulsars are shown.}
    \label{fig:post_A}
\end{figure}

The parameters and their priors are listed in Table~\ref{tab:prior}. In particular, we use uniform (instead of log-uniform) priors for the dark matter amplitudes and we justify our choices as follows. For detection purpose, if the signal amplitude can span across several orders of magnitude, the prior is often chosen to be log-uniform to yield an unbiased parameter estimation. However, for the purpose of setting upper limits, uniform priors are often used for the signal amplitude since the prior of $A$ has to be finite at $A=0$. Otherwise, if a log-uniform prior is used instead, the prior will diverge at $A=0$ and the precise value of the upper limit on $A$ will depend on the lower-cut of $A$~\cite{Arzoumanian_2016}. Considering that the data are consistent with $A=0$ (i.e. no signal), no physically motivated values can be chosen for the lower-cut of $A$, rendering such dependence undesirable. Since the main objective of this work is to place constraints of dark matter (rather than claiming detection), we use uniform priors on $A$. Note that this does skew the posterior distribution into higher values of $A$, which indicates that the upper limits that we obtain are conservative bounds. The red-noise amplitude, however, does not have such restriction. It has been shown that using uniform priors on the red noise amplitude can lead to overstated Bayesian upper limits by transferring the signal power to the red noise process~\cite{Hazboun_2020}, so we choose to use log-uniform priors instead. We show the posterior distribution of the dark matter amplitude of one of the pulsars and the maximum amplitude across the entire PTA in Fig.~\ref{fig:post_A}. 

\begin{table*}[t]
  \renewcommand{\arraystretch}{1}
  \begin{tabular}{llll}
	\hline\hline
	Parameter & Description & Prior & Comments \Tstrut\Bstrut \\ \hline
    \multicolumn{4}{c}{Red noise} \\ 
	$A_{\red}$ & Red noise power-law amplitude & Log-Uniform [$-19$, $-12$]	& one parameter per pulsar \Tstrut \\
    $\gamma_{\red}$ & Red noise power-law spectral index & Uniform [0, 7]	& one parameter per pulsar \Tstrut\Bstrut \\ \hline
    \multicolumn{4}{c}{Dark Matter} \\
    $A_{\stat}$ & Static dark matter amplitude & Uniform $\pm$[$10^{-21}$, $10^{-13}$]	& one parameter per pulsar \Tstrut \\
    $A_{\dyn}$ & Dynamic dark matter amplitude & Uniform $\pm$[$10^{-20}$, $10^{-12}$]	& one parameter per pulsar \Tstrut \\
    $t_0/T$ & Dynamic dark time of arrival & Uniform [0.1, 0.9]	& one parameter per pulsar \Tstrut \\\\
    \hline\hline
\end{tabular}
\caption{\label{tab:prior} Parameters and priors used in the mock data analysis. The notation Uniform $\pm$[$\dots$] stands for the union of Uniform [$+\dots$] and Uniform [$-\dots$]. The effects of white noise are accounted for by marginalizing over a multiplicative factor in front of the errors on the timing residuals.}
\end{table*} 

\begin{figure}[t!]
    \centering
    \includegraphics[width=\textwidth]{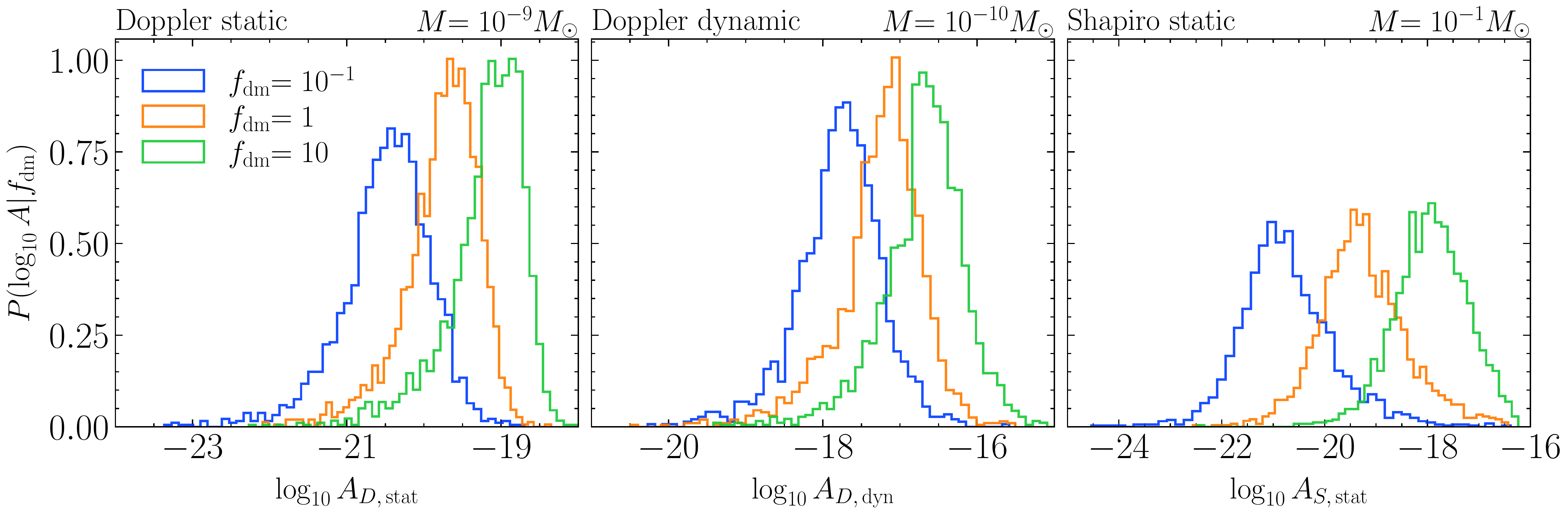}
    \caption{Conditional probability $P(\log_{10}A|f_{\dm})$ obtained by the MC for different values of $f_{\dm}$, assuming SKA parameters. The three panels correspond to Doppler static, Doppler dynamic and Shapiro static respectively.}
    \label{fig:mc_A}
\end{figure}

\begin{figure}[t!]
    \centering
    \includegraphics[width=\textwidth]{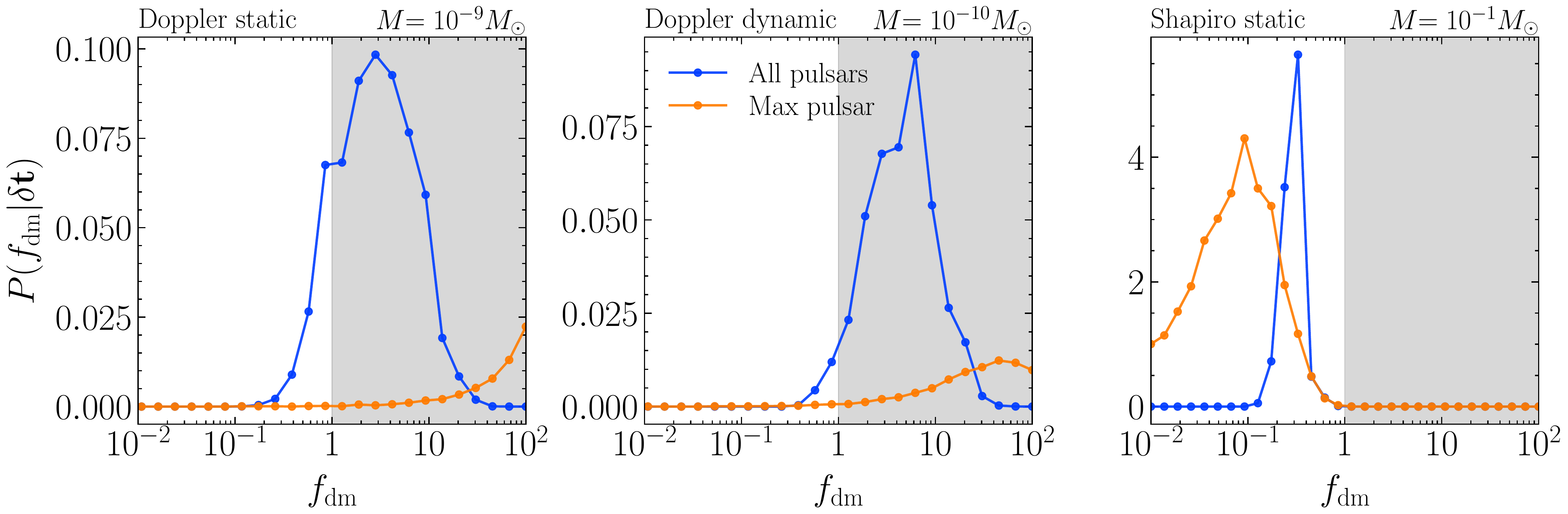}
    \caption{Posterior probability $P(f_{\dm}|\delta\vec{t})$ assuming SKA parameters and only white noise in the mock pulsars. The three panels show the results for Doppler static, Doppler dynamic and Shapiro static respectively. The lines labelled `All pulsars' use Eq.~\eqref{eq:post_prop_multi} to compute the posterior distribution, while the lines labelled `Max pulsar' use Eq.~\eqref{eq:post_prop_multi_max}.}
    \label{fig:posterior_f}
\end{figure}

\begin{figure}[t!]
    \centering
    \includegraphics[width=\textwidth]{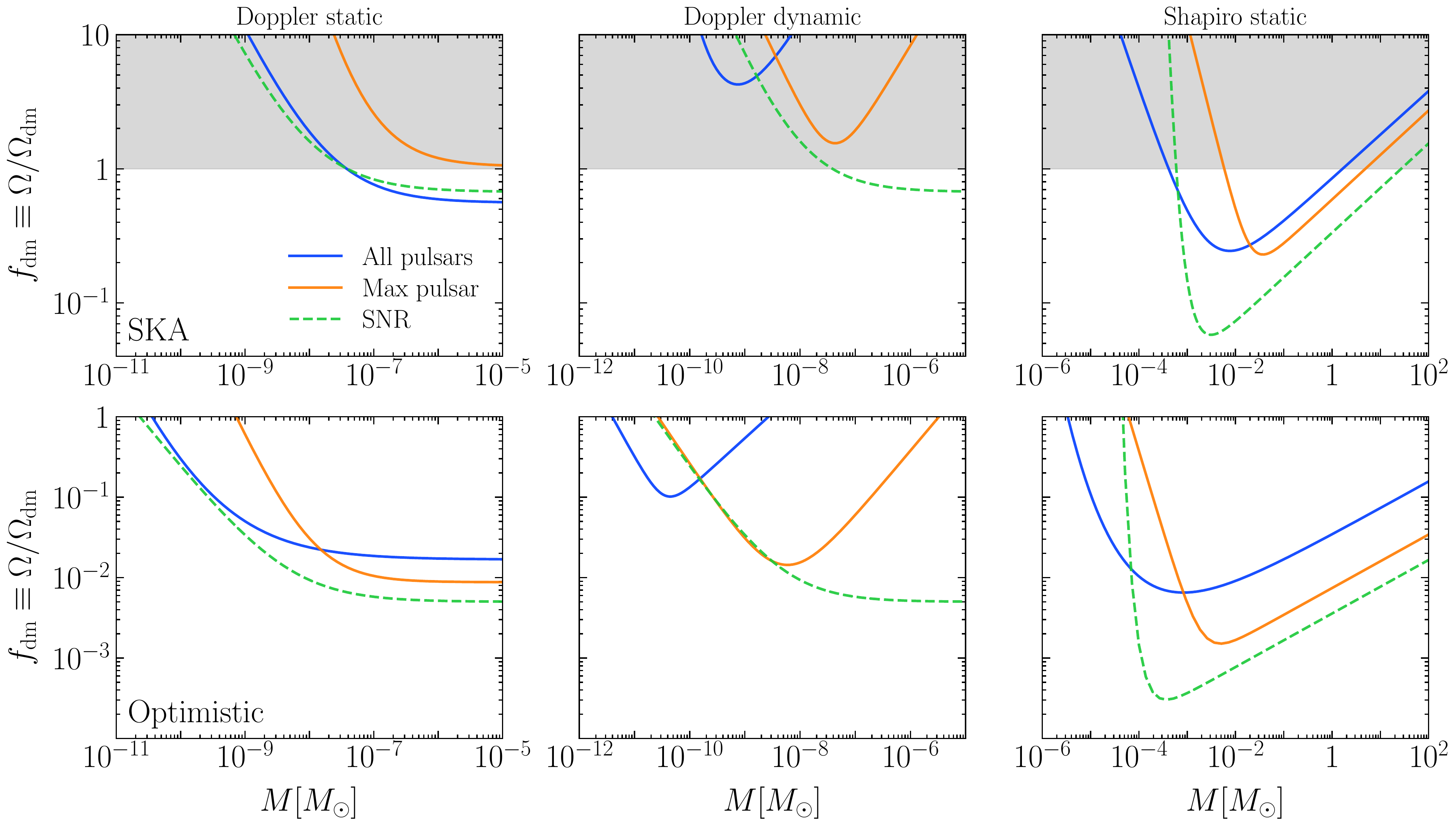}
    \caption{The 90$^\text{th}$ percentile upper limits on the PBH dark matter abundance $f_{\dm}\equiv\Omega/\Omega_{\dm}$ for different PBH masses, $M$. The top and bottom rows correspond to the SKA and optimistic parameters defined in Table~\ref{tab:pulsar_parameters}, while the three columns corresponding to the Doppler static, Doppler dynamic and Shapiro static searches, respectively. The results for this work are shown in solid lines while the dotted lines denote the projected sensitivity using the frequentist formalism developed in Refs.~\cite{Ramani_2020, Lee:2020wfn}. Note that the previous results quoted here do not distinguish between static and dynamic searches. The lines labelled `All pulsars' and `Max pulsar' labels show the upper limits derived using all pulsars and only the pulsar with maximum signal amplitude respectively.}
    \label{fig:constraints}
\end{figure}

\begin{figure}[t!]
    \centering
    \includegraphics[width=\textwidth]{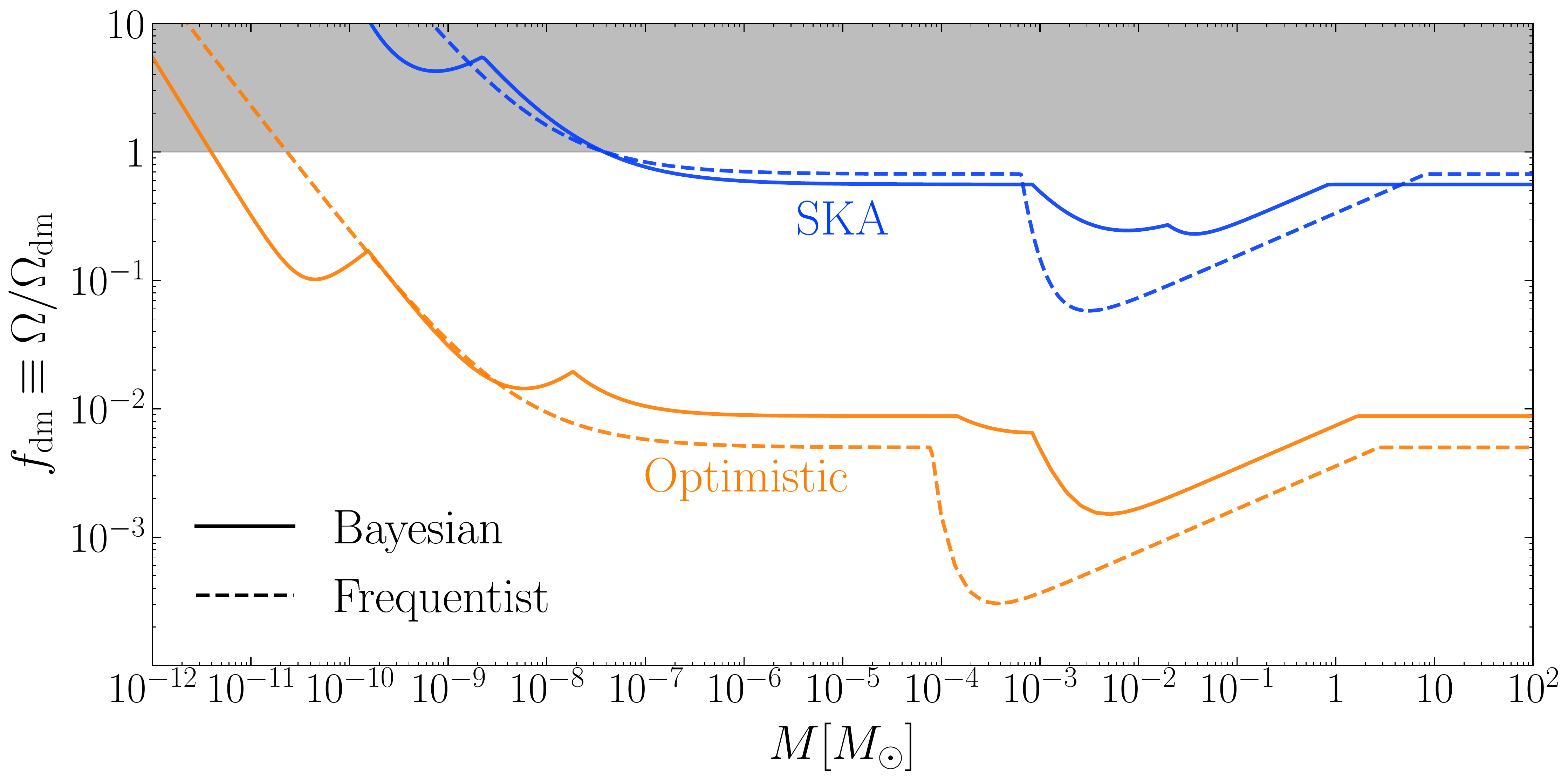}
    \caption{The most stringent 90$^\text{th}$ percentile upper limits on the PBH dark matter abundance $f_{\dm}\equiv\Omega/\Omega_{\dm}$ for different PBH masses, $M$. The results in the present work are labelled as `Bayesian' while the sensitivity projections in Refs.~\cite{Ramani_2020, Lee:2020wfn} are labelled as `Frequentist'.}
    \label{fig:best}
\end{figure}

To place an upper limit on the PBH dark matter abundance $f_{\dm}$, we first use the MC simulations described in Sec.~\ref{subsec:stat_dyn_signals} to compute the conditional probability $P(A|f_{\dm})$ for different choices of $M$ and the same pulsar parameters as the mock data. We show the probability for some choices of $f_{\dm}$ and $M$ in Fig.~\ref{fig:mc_A}. We see that higher values of $f_{\dm}$ lead to larger amplitudes. The inferred posterior distributions of $f_{\dm}$ are then computed using Eqs.~\eqref{eq:post_prop_multi}-\eqref{eq:post_prop_multi_max}, and are shown in Fig.~\ref{fig:posterior_f}.\footnote{Shaded regions in Figs.~\ref{fig:posterior_f}-\ref{fig:best} correspond to $f_{\dm} > 1$, which are unphysical if gravitation is the only interaction between the pulsar and the dark matter, but can be possible in the presence of additional forces.} By comparing Fig.~\ref{fig:post_A} and Fig.~\ref{fig:mc_A}, we observe that if $f_{\dm}$ is either too large or too small, the two probability distributions in Eqs.~\eqref{eq:post_prop_multi}-\eqref{eq:post_prop_multi_max} do not overlap at all, leading to $P(f_{\dm}|\delta \vec{t})=0$. Hence the posterior distributions of $f_{\dm}$ shrink to zero on both ends similar to the posterior distributions of $A$ in Fig.~\ref{fig:post_A}. The $p^\text{th}$ percentile upper limit constraints on $f_{\dm}$, $f_p$, are then derived by requiring $\int_0^{f_p} P(f_{\dm}|\delta \vec{t}) df_{\dm} = 100 \times p$.

Finally, the 90$^\text{th}$ percentile upper limits on $f_{\dm}$ for PBHs are shown in Fig.~\ref{fig:constraints}. These results are compared to the sensitivity projection in our previous works described in Refs.~\cite{Ramani_2020, Lee:2020wfn} using the same pulsar parameters. Our previous works use a matched-filter procedure to compute the signal-to-noise ratio (SNR) from PBHs relative to white noise, and derive the upper limits on $f_{\dm}$ by putting an appropriate cut on the SNR. We see that with both SKA and optimistic pulsar parameters, the constraints agree with each other to within a factor of two for most PBH masses. The only mass range where the results significantly differ from each other is $10^{-3}\textup{--}10^{-1}\,M_{\odot}$ (SKA) and $10^{-4}\textup{--}10^{-2}\,M_{\odot}$ (optimistic) for the Shapiro search, where our constraints on $f_{\dm}$ are weaker by around an order of magnitude. We also show the most stringent upper limits (for a given PBH mass) on $f_{\dm}$ from both the `max/all pulsar' searches in Fig.~\ref{fig:best}.

Here we summarize the differences between our previous works and this work. First, our previous work draws constraints using the SNR, which is a frequentist interpretation of the data. This work derives the constraints using the posterior distribution, which is Bayesian in nature. It is not uncommon for results from frequentist and Bayesian inferences to differ from each other by $\mathcal{O}(1)$ numbers. Second, the latest iteration of our previous work~\cite{Lee:2020wfn} does not distinguish the static and dynamic signals, because a Monte Carlo was used to generate the signals and smoothly interpolate between dynamic and static regimes. In this work, we must divide the signal into static and dynamic signals for the ease of signal parameterization in the data. This leads to a deterioration of the constraints when the mass $M$ falls under the transition region between the static and the dynamic regimes. For the Doppler case, this deterioration is not significant. However, for the Shapiro case, since we do not carry out the Shapiro dynamic analysis, the constraint is significantly weakened at the edge of the static mass regime (as commented in the previous paragraph). While this weakening is due to calculational limitations in the Bayesian analysis is unfortunate, we also note the limited utility of the Shapiro searches for even moderately lower concentration dark matter subhalos \cite{Ramani_2020}, suggesting that for a broad range of dark matter models, Doppler searches will be the dominant tool. Finally, our previous work only draws constraints using the maximum SNR among all the pulsars, while in this work we also consider the possibility of studying the dark matter signals in all the pulsars simultaneously. As indicated in Fig.~\ref{fig:constraints}, this leads to a better reach for some mass ranges (e.g. $M<10^{-2}\,M_{\odot}$ for the Shapiro signal with SKA parameters).

\subsection{Effects of Red Noise}

Realistic PTA data contain red noise. Some pulsars contain intrinsic red noise, while a stochastic GWB can also induce a red noise process correlated among all pulsars. For instance, a common red noise process with median amplitude $A=1.92\times 10^{-15}$ and spectral index $\gamma=13/3$ is reported by NANOGrav in Ref.~\cite{Arzoumanian:2020vkk}. For completeness, we briefly consider the effect of red noise, such as the SMBHB background, on a PTA's ability to detect dark matter. 

\begin{figure}[t!]
    \centering
    \includegraphics[width=0.7\textwidth]{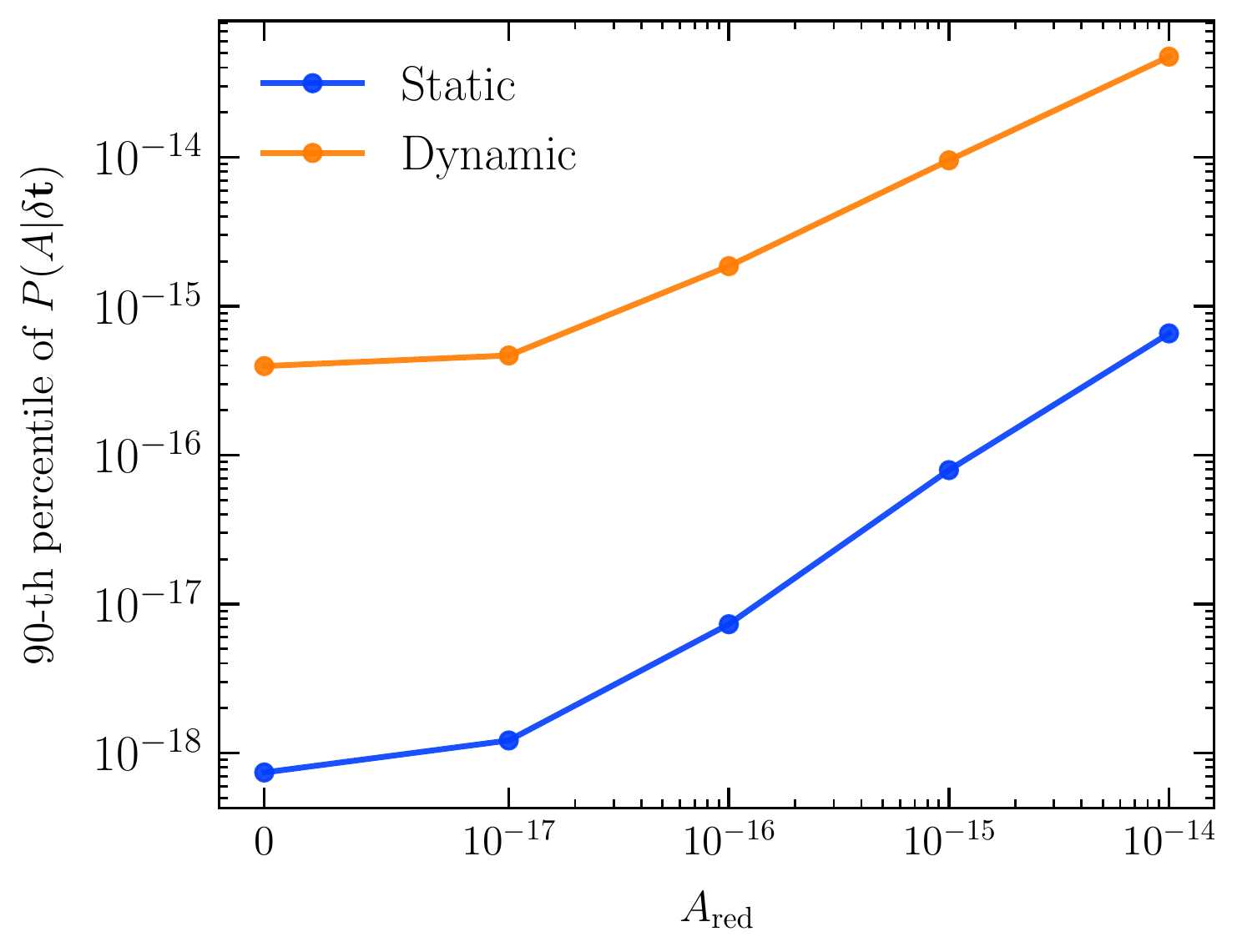}
    \caption{The $90^\text{th}$ percentile of $A_{\stat}$ and $A_{\dyn}$ in the presence of different red noise amplitudes. The data point for $A_{\mathrm{red}}=0$ corresponds to no red noise at all.}
    \label{fig:red}
\end{figure}

Instead of the upper limits on $f_{\dm}$, we report the effects of red noise on the posterior distribution of the dark matter amplitudes $A_{\stat}$ and $A_{\dyn}$ in Fig.~\ref{fig:red}. The presence of the red noise shifts the posterior distribution towards large amplitudes, implying that the constraints on the amplitudes (hence $f_{\dm}$) worsen. To quantify the effects, we show the 90$^\text{th}$ percentile of $A_{\stat}$ and $A_{\dyn}$. As shown in Fig.~\ref{fig:red}, a red noise process with $A_{\red}=10^{-15}$ would increase the upper limits on $A_{\stat}$ and $A_{\dyn}$ by 2 and 1.5 orders of magnitude respectively. The PBH dark matter abundance $f_{\dm}$ scales as $A_{D,\,\stat}$, $A_{D,\,\dyn}^2$ and $A_{S,\,\stat}^{2/3}$ respectively, meaning that, in any case, the upper limits on $f_{\dm}$ worsen by over an order of magnitude when red noise is present in the data.\footnote{In practice, instead of only considering the upper limits on $A$, one would have to perform the overlapping integrals using Eqs.~\eqref{eq:post_prop_multi}-\eqref{eq:post_prop_multi_max} to compute the posterior distribution of $f_{\dm}$. Hence this analysis is an order of magnitude estimate of the effects of $A_{\red}$ on $f_{\dm}$. We did not perform a full analysis on mock data with red noise since that would require us to run the MC simulations with unrealistically high $f_{\dm}$, which is computationally challenging.}

\section{Conclusions}
\label{sec:conclusions}

In this work, we have provided a Bayesian framework for detecting dark matter substructure with Pulsar Timing Arrays, which bridges the gap between our previous work~\cite{Dror:2019twh, Ramani_2020, Lee:2020wfn} and realistic PTA data. Using mock data with well-motivated pulsar parameters, we found that for mock pulsars with white noise only, the upper limits placed on the PBH dark matter abundance agree with our previous results up to a factor of two for all mass ranges for the Doppler search and most mass ranges for the Shapiro search. This implies that non-negligible constraints on PBHs with mass $10^{-8}\textup{--}10^{2}\,M_{\odot}$ and mass $10^{-11}\textup{--}10^{2}\,M_{\odot}$ can be placed in the next decade and the decade after respectively.

We have also investigated the effects of red noise on the sensitivity of dark matter signals, where we found that the upper limits on the PBH dark matter abundance $f_{\dm}$ weaken by over an order of magnitude when red noise from supermassive black hole binaries is introduced in the present framework. While this might eliminate any hope of detecting dark matter with PTAs in the near future, we note that significant progress is being made by the PTA community in separating signals from different physical processes. In particular, if the timing residuals due to red noise (pulsar intrinsic or pulsar correlated) are identified to high precision (instead of only the amplitude and the spectral index in frequency space), we will be able to subtract the contribution from red noise and mitigate its effects, since the dark matter signal shape studied here is not degenerate with the red noise. We hope that this work will motivate future work in the PTA community in separating signals of different sources.

This work only formulates the detection of PBHs and PBH-like substructures. To distinguish between different dark matter models, it is important to also develop a formalism that works for dark matter substructure with general halo mass functions and density profiles. In addition, many dark matter models include additional couplings between dark matter and the standard model beyond gravitational interactions. Such classes of dark matter also produce signals that can be potentially detected by PTAs~\cite{Porayko_2018}. We leave these analyses for future work.

\subsection*{Acknowledgments}

VL, TT and KZ are supported by the U.S. Department of Energy, Office of Science, Office of High Energy Physics, under Award Number DE-SC0021431 and a Simons Investigator award. SRT acknowledges support from NSF grant AST-\#2007993 and PHY-\#2020265, and a Dean's Faculty Fellowship from Vanderbilt University's College of Arts \& Science. The computations presented here were conducted on the Caltech High Performance Cluster, partially supported by a grant from the Gordon and Betty Moore Foundation.


\bibliography{bibliography}

\end{document}